# THE DEGREES OF SYMMETRY FUNCTIONS

## (preliminary version)

**Héctor Rabal, Nelly Cap**

*Centro de Investigaciones Ópticas (CCT La Plata - CIC) and UIDET Optimo, Departamento Ciencias Básicas, Facultad de Ingeniería, Universidad Nacional de la Plata, P.O. Box 3, Gonnet, La Plata 1897, Argentina.*

## Abstract

We propose a generalization of the concept of symmetry as a continuous function of the reference center or line location. We suggest that this concept can be applied to many closed systems and exploring its time evolution. When the function changes along time it is an indication that external factors are acting on the system.The change in the symmetry function along time can be thought as a current in the symmetry degree and its time behavior a measure of the effect of the external agent.

## Introduction

The surface of the Earth is in the shape of a revolution spheroid. Or, it is at least so in a first approximation. It should be wrong not to take into account that fact in its description.

It is also true that when considered in detail, the Earth shows many departures from that simplified shape and it would also be wrong not to take it into account in its description.

It would then be more proper to describe it as a revolution object with details that are not so symmetric. It would complete a more appropriate description. If the contribution of both, symmetric and non-symmetric components could be quantified the idea should be more complete. What is the proportion of each such component?

Similar comments could be done about descriptions of many objects shapes, from waveforms in optics, edges in sand grains, growing patterns in bacterial colonies, stars debris distribution after explosions (Fig. 1), etc.

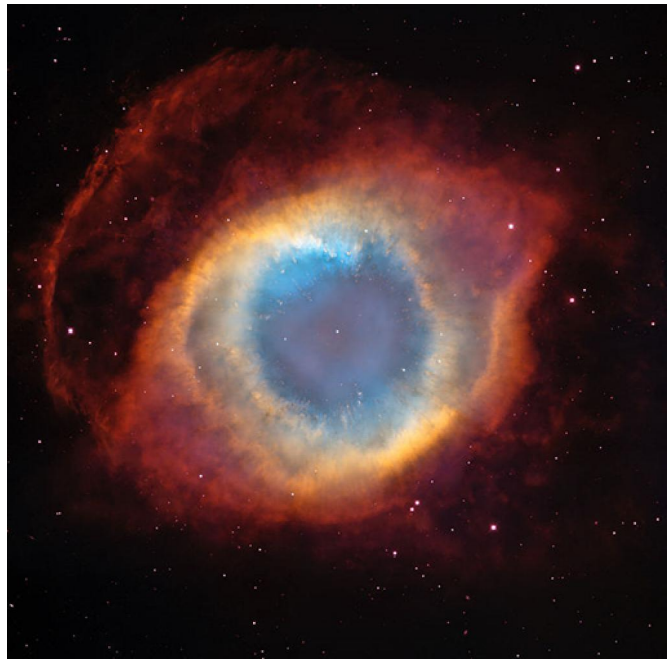

Fig. 1 Astronomic image of Helix Nebula showing approximate center symmetry (source Wikipedia).

In a paper published some years ago, one of us proposed the definition of a measure of *parity degree functions* as a property that could be demonstrated was conserved in light free propagation[1]. In this work we extend the concept to include other symmetry operations and suggest some applications. So, the lines will be quite similar because parity is a special case of symmetry, with only some modifications required for the most general case.

Symmetry is any operation that acting on an object or a function lets it in a final state that cannot be distinguished of the initial state.

Invariants play important roles in Physics and other sciences because they allow obtaining information about how a system evolves without detailed knowledge of intermediate states.

A conservation law is a statement that a physical magnitude does not change (is conserved) during interactions occurring within closed systems. The utility of the knowledge of the conservation laws of linear and angular momentum, energy, charge, etc. is well known.

In Quantum Mechanics, parity, one specific symmetry operation, is the property of a wave function that describes the behavior of the system whose physical coordinates are related by inversion about a center. If parity is even, the wave function does not change and if it is odd the wave function is only changes in sign, so that observables do not change. Other examples are found in crystallography, etc. The concept of parity usually assigns a discrete value 1 or -1 to the magnitude it acts upon.

In the usual mathematics definition of symmetry, there is no intermediate state between a symmetric function and a function not exhibiting any obvious symmetryproperty. It would be convenient if a continuous magnitude related to the concept of symmetry could be available to deal with cases where symmetry is not obvious. If this magnitude could be measured and predicted under current situations it could provide useful information.

The aim of this paper is to define a more general concept related to partial symmetry, named the Degree of Symmetry (DS) function and look for conditions for its conservation and possible applications.

It is a generalization of the parity function already defined in reference 1 that was shown to be conserved in optical waves propagation. We are going to follow similar lines.

## 2. Definition of the Degree of Symmetry (DS) function

For this definition we are going to consider a special case of symmetry, namely central symmetry, as example but the same or very similar treatment can be performed for other symmetries. The same applies for other number of dimensions.

Let us consider a complex scalar field described by a bounded function

$$f(x,y)$$

In *x,y* coordinates with respect to a certain origin O.

Then, the function can be trivially written as the sum of its (central) symmetrical part *S(f)* plus its anti-symmetrical part *A(f)*.

$$f = S + A$$

where

$$S(f) = \frac{f(x,y)+f(-x,-y)}{2} \quad (1)$$

and

$$A(f) = \frac{f(x,y)-f(-x,-y)}{2} \quad (2)$$

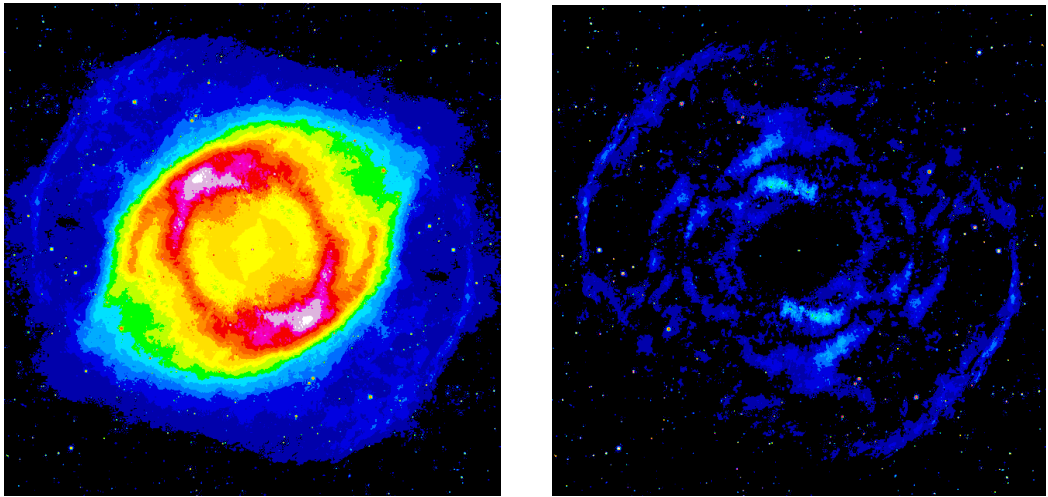

a                        b

Figure 2.The Symmetrical part a) and the Anti symmetrical part b) of Figure 1. The center of symmetry O is in the center of the Figure.

The origin O about which the symmetric and anti-symmetric parts are defined can be generalized. Symmetric (S) and Antisymmetric (A) parts with respect to *any* arbitrary origin $O´(xo, yo)$ can be written as:

$$S(f) = \frac{f(xo+x, yo+y) + f(xo-x, yo-y)}{2} \quad (3)$$

$$A(f) = \frac{f(xo+x, yo+y) - f(xo-x, yo-y)}{2} \quad (4)$$

The Degree of Symmetry *DS* of the field *f(x,y)* about the point ($x_0$, $y_0$) is defined as:

$$DS\{f\}(xo, yo) = \frac{\int (|S(f,xo,yo)|^2 dxdy)}{\int f(x,y)^2 dxdy} \quad (5)$$

where the integration interval is the support of *f(x,y)* and

$$0 < \int |f(x,y)|^2 \, dxdy < \infty \quad (6)$$

Notice that the denominator, the total energy contained in *f(x,y)*, is a constant value. Notice also that the *DS* is a number that depends on the choice of an arbitrary reference as is the location of the symmetry center (or axis, or plane or other, depending on the chosen symmetry). In some cases it is natural to choose the origin O as the mass center or any other self-evident point. Nevertheless, all the preceding description can be applied in the same way for any arbitrary choice of the reference.

When this reference location is allowed to take continuous values, the DS becomes a function.

Eq. (5) can be written, by using eq. (4) and the convolution operator in two dimensions, indicated as ⊗,

$$DS\{f\}(xo, yo) = \frac{1}{2}\left\{1 + \frac{(f \otimes \otimes f^*)}{\int f(x,y)^2 dxdy}(2xo, 2yo)\right\} \quad (7)$$

where * indicates complex conjugation.

After eq. (7), but also from its definition, it can be seen that if *f(x,y)* has compact support, then the *DS* consists of an inconsequential constant additive term ½ plus a bounded function with compact support.

The *DS* is then a function generally smoother than (or, eventually as smooth as) the function itself.

If one (or both) of the integrals in (5) is divergent, the DS is defined as

$$DS\{f\}(xo, yo) = \frac{1}{2}\left\{1 + \lim_{A,B\to\infty} \frac{\int_{-A}^{A}\int_{-B}^{B} f(xo+x,yo+y)f^*(xo-x,yo-y)dxdy}{\int_{-A}^{A}\int_{-B}^{B}|f(x,y)|^2 dxdy}\right\} \quad (8)$$

$$\int |f(x,y)|^2 dx\, dy \neq 0$$

If both integrals in eq. (8) converge, then eq. (8) and eq. (5) coincide.

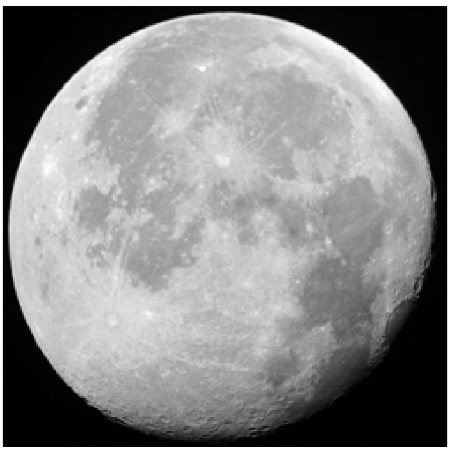
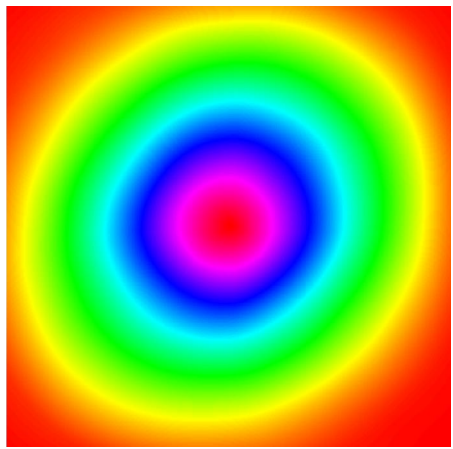

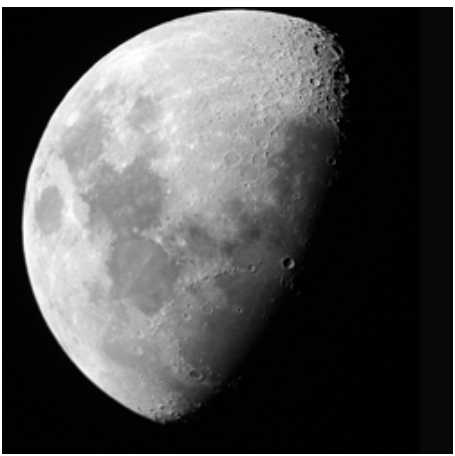
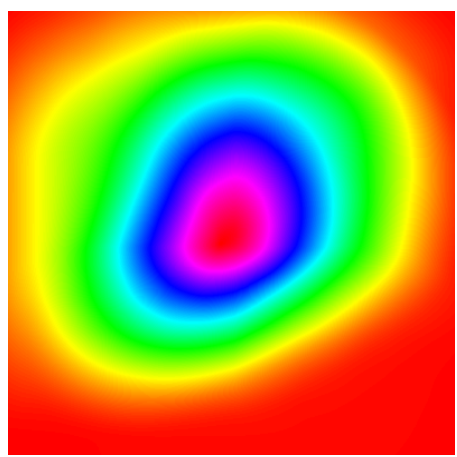

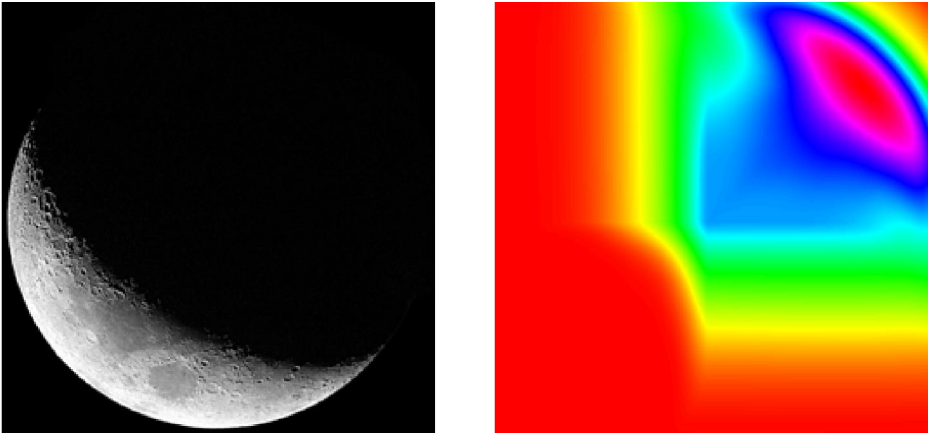

Figure 3 Three images of the Moon at left and their *DS* functions (with respect to central symmetry) in false color at right.

## 3. Properties of the DS function

**Property 1.** The DS function does not change under complex conjugation, as can be seen from eq. (5)

$$DS\{f^*\}(xo, yo) = DS\{f\}(xo, yo) \quad (9)$$

**Property 2.** The DS function does not change under multiplication of the function by a complex constant, as can be seen from eq. (6). If $c \in C$

$$DS\{f\}(xo, yo) = DS\{cf\}(xo, yo) \quad (10)$$

(and then, under arbitrary phase changes exp[i$\varphi$] with $\varphi$ a constant)

**Property 3.** If the Degree of Anti Symmetry *DAS* function of the scalar function f at point $x_0, y_0$ is defined as

$$DAS\{f\}(xo, yo) = \frac{\int(|A(f,xo,yo)|^2 dxdy)}{\int f(x,y)^2 dxdy} \quad (11)$$

Then, it is verified that:

$$DS\{f\}(xo, yo) + DAS\{f\}(xo, yo) = 1 \quad (12)$$

**Property 4** As a consequence of eq. (12) the *DS* is bounded to the [0,1] interval, it is

$$DS\{f\}(xo, yo) \leq 1 \quad (13)$$

**Property 5** If *f(x,y)* is symmetric or anti symmetric with respect to (x,y) = (0,0) then *DS {f}(xo, yo)* is symmetric with respect to (0,0) as can be seen from Eqs. (4) and (5).

**Property 6** If f(x,y) is symmetric with respect to (x,y) = (0,0), then

$$DS\{f\}(0,0) = 1 \quad (14)$$

If f(x,y) is anti-symmetric with respect to (x,y) = (0,0), then

$$DS\{f\}(0,0) = 0 \quad (15)$$

So, if a symmetric function has constant *DS* then its *DS* is 1 and if an anti-symmetric function has constant *DS* then its *DS* is 0.

**Property 7** For all $a \in R$

$$DS\{f(ax, ay)\}(x_0, y_0) = DS\{f(x, y)\}(ax_0, ay_0) \quad (16)$$

**Property 8** If f is such that

$$\int |f(x, y)|^2 dx\, dy < \infty$$

then

$$DS\{DS\{f\}(xo, yo)\} = 1 \quad (17)$$

**Property 9** If we assume *f* with

$$\int |f(x, y)|^2 dx\, dy > 0$$

then, if *DS(0,0)* = constant, *f* has not bounded support.

**Postulate:**

*If a system is closed, the fraction of total energy of the function describing it contained in its symmetric part, named Degree of Symmetry DS (of the chosen symmetry), does not change.*

If this DS changes it is because there is acting an external factor that produces transference of that fraction of energy between the *DS* and the *DAS*.

The converse needs not to be true. There may be external acting agents not producing *DS* transfer.

If the effect of the external factor along time is considered, the transfer can be thought as a DS current *I* and the change in *DS* along time could eventually permit to infer the presence of the external factor.

For example, by taking images of exploding stars and measuring *DS* with respect to some symmetries and verifying its constancy or not, the presence of non-detected external factors could eventually be inferred.

*DS* current measurement along time would permit, perhaps, to obtain some information about the external agent (its location, for example).

**Example**

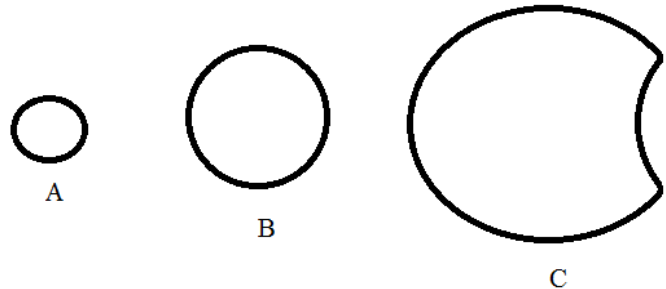

Figure 4

Figure 4 could show a scheme of the growing of a bacteria colony as observed by a certain method (conventional microscope or speckle activity images[2])

In Figure 4a) circular symmetry is perfect. Its *DS* with respect to the center is 1. The same happens in 4b). The colony grew conserving its original circular symmetry.

In 4c), it is evident that the DS is not the same. Something external has changed that degree of symmetry (the presence of other microorganism that the microscope is not able to resolve, for example).This example was chosen extreme for illustration.

An important consequence of the postulate is that it is not necessary that the two first situations should be symmetric. Any shape could be adequate for this experiment.It is enough to calculate (measure) its DS even if the image is irregular or seemingly random.The DS should be constant along time if there are not external factors acting.If the DS changes then it is an indication of the presence of external factors.

Notice that in Figure 3, the same object (the Moon) exhibits different *DS* along time, thus indicating that the system Moon-Observer is not a closed one.

The concept could be applied to (truncated) 1 D signals, for example (ECG, EEG, dynamic speckle signals, etc.) by using degree of parity as the symmetry and the fact that the choice of the origin is arbitrary.

**Theorem**

The necessary and sufficient condition for two functions f and g to have the same DS is (for central symmetry):

$$F(x,y)F^*(-x,-y) = G(x,y)G^*(-x,-y) \quad (18)$$

where F and G are the Fourier Transforms of *f* and *g* respectively, assuming that both Fourier Transforms exist.

From Eq.(7) it follows that the necessary and sufficient condition for the two functions to have the same DS is

$$f \otimes\otimes f^* = g \otimes\otimes g^* \quad (19)$$

Then, by applying the Fourier Transform, Eq. (18) is obtained.

**Conservation of the *DS* in free space propagation**

Let us consider free propagation of scalar fields

$$f(x, y; z)$$

along the (mean) direction of the coordinate axis z, with

$$\int |f(x,y)|^2 dx\, dy = 1$$

Propagation satisfies the Helmholtz equation and is a Linear Space Invariant phenomenon[3]. As such it is characterized by a transfer function. Let G be the Fourier Transform (FT) of the propagated field *g* due to a pupil described by the function *f*, and let F be the Fourier Transform of *f*. Then G, the Fourier Transform of the propagated field *g* is related to the FT F of the field *f* before propagation by the transfer function of the wave propagation[3]

$$H(f_x, f_y) = \begin{cases} \exp\left[i\frac{2\pi}{\lambda}\sqrt{1 - (\lambda(f_x))^2 - (\lambda(f_y))^2}\right] & \lambda|f_x, f_y| < 1, \\ 0, & elsewhere \end{cases}$$

(19)

as

$$G = HF \qquad (20)$$

with z the propagation distance, $\lambda$ the wavelength of the light and *fx, fy* the associated spatial frequencies.

Then, the *DS(f)* is the same as *DS(g)*.

**Corollary (Conservation of the DS)**

If the function *g* is the function describing the optical field *f* after free propagation a distance z, then

$$DS_g(x_0, y_0) = DS_f(x_0, y_0)$$

The relationship between the functions F(x) and G(x) is [ref. 2]

$$G(f_x, f_y) = \exp\left[i\frac{2\pi}{\lambda}\sqrt{1 - (\lambda(f_x))^2 - (\lambda(f_y))^2}\right] F(f_x, f_y)$$

Then Eq. (18) is verified.

This result implies that any Linear Shift Invariant (LSI) System conserves the DS.

**If a function *f* is used as input of a system *S* and *g* is the output, then DS(*f*) = DS (*g*) if and only if system S is a LSI system.**

Optical waves propagation is a LSI but for Fraunhofer regime, where shift invariance is broken.

Physically, the conservation of the DS function is a consequence of the conservation of energy and indicates that the energies in the symmetric and anti-symmetric part of the function are independently conserved after free propagation. Interference between these parts contains no total energy.

**Conclusions**

We have defined a magnitude, the DS corresponding to a symmetry, that can be associated to scalar fields with great generality and can be used for many different symmetries. It was found that this magnitude, generalized with respect to the reference location, is conserved in LSI systems, such as light propagation. This necessary and sufficient condition only exists in Nature as an approximation. It could be concluded that the same occurs with DS conservation.

DS eventual change could be an indication that the system is not closed and some external agent is acting on it. In spite of the fact that no physical system is rigorously Linear Shift Invariant, LSI systems are a commonplace in many fields, as for example, in optics, signals processing and electronic engineering because many of them can be safely approximated in this way for adequatelysmall ranges.

The conservation of DS could be used, for example, as a tool for the stabilization of a LSI system in front of possible external interferences. If it is detected that the DS changes, then an automatic control could be programmed to introduce changes in the system parameters to restore the original DS and, with it, the (approximate) LSI. This is a continuous generalization of the usual parity check in computational science.

## Acknowledgements

This work was supported by a Grant PICT 2008-1430 from the Agencia Nacional de Promoción en Ciencia y Tecnología (ANPCyT),by Consejo Nacional de Investigaciones Científicas y Técnicas (CONICET), Comisión de Investigaciones Científicas de la Provincia de Buenos Aires, and by Universidad Nacional de La Plata (UNLP), Argentina.

The pictures of the Moon were kindly provided by Elvio Alanís, Observatorio UNSa.

To Beatriz Ruiz *(In memoriam).*